\documentclass[12pt,aps,preprint,nofootinbib]{revtex4}

\usepackage{graphicx}
\usepackage{cancel}
\usepackage{amssymb}
\usepackage{textcomp}
\usepackage{amsmath}
\usepackage{bm}
\usepackage{times}
\usepackage{epsfig}
\usepackage{epstopdf}
\usepackage{color}
\usepackage{multirow}

\def\lsim{\mathrel{\raise.3ex\hbox{$<$\kern-.75em\lower1ex\hbox{$\sim$}}}}
\def\gsim{\mathrel{\raise.3ex\hbox{$>$\kern-.75em\lower1ex\hbox{$\sim$}}}}

\def\beq{\begin{equation}}
\def\eeq{\end{equation}}
\def\be{\begin{equation}}
\def\ee{\end{equation}}
\def\bea{\begin{eqnarray}}
\def\eea{\end{eqnarray}}

\def\to{\rightarrow}

\begin{document}

\title{Probing NLSP Top Squark from Heavy Gluino at the LHC}

\author{Tong Li}
%\author{Tong Li}
%\email{tli@udel.edu}

\affiliation{
Bartol Research Institute, Department of Physics and Astronomy,
University of Delaware, Newark, DE 19716, USA
}

\begin{abstract}
In this work we explore the 7 TeV LHC implication on light top squark (stop) and investigate discovery limits at both 8 and 14 TeV LHC using same-sign dilepton signature.
We consider a simplified spectrum with gluino pair production followed by heavy gluino decay into on-shell stop and top quark in pair. The light stop is then assumed to uniquely decay into $c\tilde{\chi}_1^0$, $bW^+\tilde{\chi}_1^0$ or $t\tilde{\chi}_1^0$.
The same-sign dilepton search at 7 TeV LHC places a lower bound of about 800 GeV on the gluino mass in this scenario.
We find that the 8 TeV LHC with 30 fb$^{-1}$ luminosity can reach 960 (900) GeV, 1.2 (1.0) TeV and 1.3 (1.1) TeV of gluino mass for stop decay product of $c\tilde{\chi}_1^0$, $bW^+\tilde{\chi}_1^0$ and $t\tilde{\chi}_1^0$ respectively, corresponding to $3 (5) \sigma$ significance. The discovery limits of heavy gluino mass at the 14 TeV LHC with 100 fb$^{-1}$ luminosity are 1.45 (1.35) TeV, 1.6 (1.55) TeV and 1.9 (1.68) TeV.
% for c, bW and t mode respectively with significance $S/\sqrt{B}>3 (5) \sigma$.
\end{abstract}

\maketitle

%%%%%%%%%%%%%%%%%%%%%%%%%%%%%%%%%%%%%%%%%%%%%%%%%%%%%%%%%%%
\section{Introduction}
%%%%%%%%%%%%%%%%%%%%%%%%%%%%%%%%%%%%%%%%%%%%%%%%%%%%%%%%%%%

Low scale supersymmetry (SUSY), augmented by an unbroken R-parity, largely overcomes the gauge hierarchy problem encountered in
the Standard Model (SM) and also provides a compelling cold dark matter candidate.
%The reason of expecting supersymmetry to be
%discovered at a relatively low energy is that people postulated supersymmetry to provide a natural explanation of the mass
%scale of electroweak symmetry breaking.
In most supersymmetric models, like mSUGRA/constrained minimal supersymmetric model (CMSSM)~\cite{cmssm},
the SUSY mass scale is very closely tied to the mass scale of the Higgs vacuum expectation value and all SUSY mass terms are of the order of a few
hundred GeV. Thus the first two family squarks and gluino have very large production cross sections at hadron colliders. However, these predicted large
cross sections have not been discovered with multiple energetic jets plus large missing energy signature with the integrated luminosity of up to 5 fb$^{-1}$ at $\sqrt{s}=7$ TeV LHC~\cite{atlasjets,cmsjets}.

Despite the lack of evidence for SUSY at the LHC so far, SUSY remains a very attractive possibility for physics at the TeV scale. Cohen, Kaplan and Nelson once
introduced the ``more minimal supersymmetric Standard Model''~\cite{mmssm}, in which only the third generation sfermions and gauginos are light.
%, and the first two family squarks and sleptons are very heavy.
This scenario of superparticle spectrum can also be realized in a class of supersymmetric models accommodating the stable lightest neutralino (LSP) compatible with the WMAP dark matter measurement~\cite{lsp,wmap}.
%Generally, the annihilation cross section of a pure bino LSP with mass of around 100 GeV is small and does not permit one to easily reproduce the required relic
%dark matter abundance~\cite{relicabundance}. An interesting scenario which enhances the bino annihilation cross section is the co-annihilation of bino and the
%lighter top squark (stop).
In this scenario the bino LSP and the relevant lightest squark, namely NLSP stop (where NLSP stands for next-to-lightest supersymmetric particle) are sufficiently close together in mass, such that the ensuing co-annihilation processes in the early universe allow one to reproduce the desired
bino relic density~\cite{relicabundance}. Besides the co-annihilation scenario, there are other motivations supporting the lighter stop to be the NLSP with the mass even smaller than the top quark mass, for instance the electroweak baryongenesis~\cite{carlos} and the naturalness in electroweak symmetry breaking in the MSSM~\cite{naturalness}.

%For instance, in the framework of MSSM, electroweak baryongenesis would be possible in the presence of light stop and thus favor $M_{\tilde{t}_1}, M_h\lesssim %130$ GeV~\cite{carlos}. Also, the stop-top loop diagrams contribute to the $m_{H_u}^2$ parameter and the naturalness in electroweak symmetry breaking in the %MSSM consequently requires the upper bound on the lighter stop mass to be around a few hundred GeV~\cite{naturalness}.
%Recently the NLSP stop solutions as the low
%energy consequence of implementing $b-\tau$ Yukawa unification in the mSUGRA/CMSSM have been explored~\cite{btau}.

The direct search for NLSP stop, especially in the region of nearly degenerate stop and LSP neutralino masses, is challenging and has been implemented by both LEP and Tevatron~\cite{expstop1,expstop2,expstop3}.
%assuming the loop-induced NLSP stop two-body decay into a charm quark and a neutralino being dominant~\cite{kobayashi}.
They put limit on the NLSP stop mass as $M_{\tilde{t}_1}>100$ GeV from LEP-II and $M_{\tilde{t}_1}>180$ GeV from CDF Run-II.
However,
%the Tevatron has no sensitivity in stop searches for the mass difference of stop and LSP neutralino below 40 GeV, thus
the Tevatron bound does not cover the co-annihilation region above the LEP limit~\cite{expstop3}. Two alternative search methods have been proposed to detect light stop instead of searching
for pure stop pair production events. One of them is to consider stop pair production associated with a hard QCD jet~\cite{carlos2}. In the co-annihilation region, there will be minimal hadronic activity associated with the stop decay and therefore this channel would effectively lead to events with hard jets and large missing energy. Recently, it is found that co-annihilation region with $M_{\tilde{t}_1}\lesssim 140$ GeV is essentially ruled out in light of 7 TeV LHC data with 1 fb$^{-1}$ integrated luminosity~\cite{adeel,bin}. The other proposed method takes advantage of the Majorana fermion feature of gluino and considers gluino pair production followed by gluino decay into on-shell stop and top quark~\cite{kraml,martin}. The pair production of gluinos
leads to the events containing same-sign dilepton signature arising from the same-sign top quarks leptonic decay and induces negligible Standard Model (SM) backgrounds.

On the other hand, LHC collaborations, ATLAS and CMS have performed analyses for events containing two same-sign isolated leptons plus hadronic jets and missing energy at a center-of-mass energy of 7 TeV~\cite{atlasss2,cmsss5}.
%In particular, CMS recently analyzes a search for anomalous production in the same-sign dilepton final state with at least two $b$-jets and missing energy~\cite{cmsssb5}.
Good agreement was observed between the number of events in data and the SM predictions. They could apply for the second NLSP stop search scenario mentioned above, namely gluino pair production with gluino decay into top and stop and make improved constraints on relevant parameter space beyond LEP and Tevatron. In this paper, we would like to study the 7 TeV LHC impact on the scenario of heavy gluino decay into NLSP stop and predict the discovery limits at both 8 and 14 TeV LHC.

This paper is organized as follows. In section II we discuss the NLSP stop decay and production modes and outline the selection requirements employed by the LHC
collaborations. The kinematic features of NLSP stop production are also presented together with constrained parameter space in terms of the masses of gluino, stop and neutralino. Our conclusions are summarized in section III.

%%%%%%%%%%%%%%%%%%%%%%%%%%%%%%%%%%%%%%%%%%%%%%%%%%%%%%%%%%%
\section{Heavy Gluino and NLSP top squark at the LHC}
%%%%%%%%%%%%%%%%%%%%%%%%%%%%%%%%%%%%%%%%%%%%%%%%%%%%%%%%%%%

%%%%%%%%%%%%%%%%%%%%%%%%%%%%%%%%%%%%%%%%%%%%%%%%%%%%%%%%%%%
\subsection{Heavy gluino production and NLSP stop decay modes}
%%%%%%%%%%%%%%%%%%%%%%%%%%%%%%%%%%%%%%%%%%%%%%%%%%%%%%%%%%%
To explore common features of NLSP stop scenario model-independently, we adopt the concept of so-called ``simplified models'' paradigm~\cite{simmodel1,simmodel2,simmodel3,simmodel4}. Simplified models parameterize the new physics by a simple particle spectrum, its production mode and decay topologies with the masses, cross sections and branching ratios taken as free parameters. 
%They capture genetic kinematic properties of models that are
%relevant for early searches. 
Particles that are not involved in a specific signature are assumed to be decoupled. For our case,
%we will consider gluino pair production with 100\% gluino decay into top and stop. Totally
we have three free mass parameters, namely gluino mass $M_{\tilde{g}}$, the lighter stop mass $M_{\tilde{t}_1}$ and neutralino LSP mass $M_{\tilde{\chi}_1^0}$ under the assumption that all other superparticles decouple.

In the framework of MSSM with gravity mediated supersymmetry breaking, the NLSP stop $\tilde{t}_1$, with LSP neutralino, has the following decay channels
\begin{eqnarray}
\tilde{t}_1\to c\tilde{\chi}_1^0, f\bar{f}'b\tilde{\chi}_1^0, bW^+\tilde{\chi}_1^0, t\tilde{\chi}_1^0,
\end{eqnarray}
assuming other superparticles decouple in simplified model, where $f$ and $f$' stand for SM leptons or quarks. These decays are all generated at tree level except for the first channel, which is loop-induced and proceeds through off-diagonal elements of CKM matrix. The three tree level channels gradually come into play from left to right, corresponding to increasing $\Delta M\equiv M_{\tilde{t}_1}-M_{\tilde{\chi}_1^0}$. They proceed through both off-shell top quark and $W$ boson exchange (or sbottom, slepton, sneutrino, chargino), only off-shell top quark (or sbottom, chargino), and via on-shell top quark respectively. In particular, for an extremely small $\Delta M$, the NLSP stop decay products from the 4-body channel ($f\bar{f}'b\tilde{\chi}_1^0$) are much softer and thus harder to detect compared to the 2-body channel ($c\tilde{\chi}_1^0$), and have not been searched in experiments so far. Therefore, we scan the parameter region of $M_{\tilde{t}_1}$ and $M_{\tilde{\chi}_1^0}$ with varying $\Delta M$ and consider the related unique NLSP stop decay channels
\begin{eqnarray}
&&{\rm c \ mode: \ }\tilde{t}_1\to c \tilde{\chi}_1^0, \ \ m_c<\Delta M<M_W+m_b,\\
&&{\rm bW \ mode: \ }\tilde{t}_1\to bW^+\tilde{\chi}_1^0, \ \ M_W+m_b<\Delta M<m_t,\\
&&{\rm t \ mode: \ }\tilde{t}_1\to t\tilde{\chi}_1^0, \ \ \Delta M>m_t,
\end{eqnarray}
with 100\% of decay branching fraction each.

We further assume the NLSP stop associated with a top quark arises from the heavier gluino pair production and decay, with 50\% of $\tilde{g}\to t\tilde{t}_1^\ast$ or $\tilde{g}\to \bar{t}\tilde{t}_1$ decay branching ratio and mass range $M_{\tilde{g}}> M_{\tilde{t}_1}+m_t$. From the well-known
relation between gaugino masses at low energy in mSUGRA/CMSSM, namely $M_3 :
M_2 : M_1 \approx 6 : 2 : 1$, which follows from the assumption of universal gaugino masses at
high scale, the gluino mass is taken to be $M_{\tilde{g}}=6M_{\tilde{\chi}_1^0}$ in the following studies.
Together with the three NLSP stop decay modes,
one has the following same-sign dilepton productions
\begin{eqnarray}
&&\tilde{g}\tilde{g}\to (t\tilde{t}^\ast_1+\bar{t}\tilde{t}_1)(t\tilde{t}^\ast_1+\bar{t}\tilde{t}_1) \\
&&\underrightarrow{c} \ t\tilde{t}^\ast_1t\tilde{t}^\ast_1+c.c.\to bW^+\bar{c}\tilde{\chi}_1^0bW^+\bar{c}\tilde{\chi}_1^0+c.c.\to \ell^\pm\ell^\pm jjbb\cancel{E}
\label{c}\\
&&\underrightarrow{bW} \ t\tilde{t}^\ast_1t\tilde{t}^\ast_1+t\tilde{t}^\ast_1\bar{t}\tilde{t}_1+c.c.\to bW^+\bar{b}W^-\tilde{\chi}_1^0bW^+\bar{b}W^-\tilde{\chi}_1^0+bW^+\bar{b}W^-\tilde{\chi}_1^0\bar{b}W^-bW^+\tilde{\chi}_1^0+c.c.\nonumber \\
&&\to jjbbbb\ell^\pm\ell^\pm \cancel{E}
\label{bW}\\
&&\underrightarrow{t} \ t\tilde{t}^\ast_1t\tilde{t}^\ast_1+t\tilde{t}^\ast_1\bar{t}\tilde{t}_1+c.c.\to bW^+\bar{t}\tilde{\chi}_1^0bW^+\bar{t}\tilde{\chi}_1^0+bW^+\bar{t}\tilde{\chi}_1^0\bar{b}W^-t\tilde{\chi}_1^0+c.c.\to jjbbbb\ell^\pm\ell^\pm \cancel{E}.
\label{t}
\end{eqnarray}
Note that the c mode provides the same-sign dilepton signal only when same-sign tops are produced as shown in Eq.~(\ref{c}), as the stop has no leptonic decay product. By contrast, any $t$ and $\tilde{t}_1$ production configuration from gluino pair decay contributes to the same-sign dilepton signal for both the bW and t modes as shown in Eqs.~(\ref{bW}) and (\ref{t}).
We use Pythia to generate gluino pair production events, and include decay, parton showering and hadronization~\cite{pythia}. PGS-4 is used to simulate the important detector effects with CMS parameters~\cite{pgs}. The cross sections of gluino pair production are normalized to the next-to-leading order output of Prospino 2.1~\cite{prospino}.

%%%%%%%%%%%%%%%%%%%%%%%%%%%%%%%%%%%%%%%%%%%%%%%%%%%%%%%%%%%
\subsection{LHC constraint on NLSP stop from heavy gluino}
%%%%%%%%%%%%%%%%%%%%%%%%%%%%%%%%%%%%%%%%%%%%%%%%%%%%%%%%%%%
The ATLAS and CMS collaborations have reported data in terms of events containing same-sign dilepton signature in $\sqrt{s}=7$ TeV proton-proton collisions.
No excess above the SM background expectation was observed. This data can be employed, as we show below, to find more stringent constraints on the NLSP stop scenario.

In the CMS analyses corresponding to 1 fb$^{-1}$ integrated luminosity, the events considered for search regions are all required to have two leptons with the same charge, at least two jets, and $\cancel{E}_T>30$ GeV~\cite{cmsss5}. The CMS selections are defined by three baselines: events with $\mu\mu,ee,\mu e$ dilepton candidates (inclusive dileptons), events with $\mu\mu,ee,\mu e$ dilepton candidates with both leptons having $p_T>10$ GeV, at least one lepton having $p_T>20$ GeV(high-$p_T$ dileptons), and events with $\tau\tau,e\tau,\mu\tau$ dilepton candidates ($\tau$ dileptons). The following inclusive search regions are also defined to constrain the baseline selection categories: high-$H_T$ high-$\cancel{E}_T$ with $H_T>400$ GeV and $\cancel{E}_T>120$ GeV, medium-$H_T$ high-$\cancel{E}_T$ with $H_T>200$ GeV and $\cancel{E}_T>120$ GeV, high-$H_T$ low-$\cancel{E}_T$ with $H_T>400$ GeV and $\cancel{E}_T>50$ GeV, and low-$H_T$ high-$\cancel{E}_T$ with $H_T>80$ GeV and $\cancel{E}_T>100$ GeV, as shown in Table~\ref{cuts4}. The observed upper limits on events from new physics for baseline 1 with regions 1-3 (I1-I3), baseline 2 with regions 1-4 (H1-H4), and baseline 3 with region 1 (T1) are represented in the last row of Table~\ref{cuts4}.
%The observed upper limits on events from new physics are represented in Table~\ref{cuts4}.

\begin{table}[h]
\begin{center}
\begin{tabular}[t]{|c|c|c|c|c|c|c|c|c|}
  \hline
 & I1 & I2 & I3 & H1 & H2 & H3 & H4 & T1\\
  \hline
%  all (NLSP $\tilde{g}$) & 5420(3832) & 5420(3832) & 5420(3832) & 5420(3832) & 5420(3832) & %5420(3832)\\
  %\hline
  $H_T$ (GeV)& $> 400$ & $> 400$ & $> 200$ & $> 400$ & $>400$ & $>200$ & $>80$ & $>400$\\
  \hline
  $\cancel{E}_T$ (GeV)& $> 120$ & $>50$ & $> 120$ & $> 120$ & $>50$ & $>120$ & $>100$ & $>120$\\
  \hline
  2 leptons $p_T>10$ GeV & &  & & $\surd$ & $\surd$ & $\surd$ & $\surd$ &\\
  %  \hline
  $\geq 1$ lepton $p_T>20$ GeV & &  &  &  &&&&\\
%  \hline
% Leading jet $p_T$ (GeV) & $>130$ & $>130$ & $>130$ & $>130$\\
%  \hline
% Second jet $p_T$ (GeV) & $>40$ & $>40$ & $>40$ & $>40$ \\
 % \hline
% Third jet $p_T$ (GeV) & $-$ & $>40$ & $>40$ & $>40$ \\
 % \hline
% Fourth jet $p_T$ (GeV) & $-$ & $-$ & $>40$ & $>40$ \\
   \hline
  CMS $N_{{\rm exp}}$ & $3.7$ & $8.9$ & $7.3$ & $3.0$ & $7.5$ & $5.2$ & $6.0$ & $5.8$\\
  \hline
\end{tabular}
\end{center}
\caption{Summary of selection cuts and 95$\%$ C.L. upper limits on event number for signal region I1-I3, H1-H4 and T1 containing final states with same-sign isolated dilepton, jets and missing energy with 0.98 fb$^{-1}$ luminosity, following the data analyses of CMS~\cite{cmsss5}.}
\label{cuts4}
\end{table}

We apply $\sigma\times {\rm acceptance}>\sigma_{{\rm exp}}$ as exclusion requirement for each spectrum configuration, where $\sigma$ is the relevant total cross section and the acceptance is the ratio of signal events after and before selection cuts which reflects the effects of experimental efficiency.
Note that in order to avoid the possibly large uncertainty from the calibration of hadronic $\tau$, in the following analyses we do not include the region T1.
We find the region of gluino $M_{\tilde{g}}\lesssim 750$ GeV (840 GeV) for c (bW and t) mode is essentially excluded by 7 TeV LHC with the integrated luminosity of 1 fb$^{-1}$. In addition, CMS has recently released results of same-sign dilepton plus at least two $b$-jets tagged with the integrated luminosity of 4.98 fb$^{-1}$~\cite{cmsssb5}. This search is more dedicated for probing the model with NLSP stop and top quark in final state. Demanding the gluino decay to a top-stop pair and $\tilde{t}_1\to t\tilde{\chi}_1^0$ (t mode), the lower limit of gluino is set around $800-850$ GeV according to CMS analysis.

%%%%%%%%%%%%%%%%%%%%%%%%%%%%%%%%%%%%%%%%%%%%%%%%%%%%%%%%%%%
\subsection{Searching heavy gluino and NLSP stop from same-sign dilepton at 8 and 14 TeV LHC}
%%%%%%%%%%%%%%%%%%%%%%%%%%%%%%%%%%%%%%%%%%%%%%%%%%%%%%%%%%%

To search the scenario of NLSP stop from heavy gluino produciton in terms of same-sign dilepton at the LHC,
we employ the same object requirements as applied in CMS detector for the event selection~\cite{cmsssb5}
\begin{itemize}
\item two leptons of the same sign with $p_T>20$ GeV and $|\eta|<2.4$,
\item jets with $p_T>40$ GeV and $|\eta|<2.5$.
\end{itemize}
We use the default value of b-tagging efficiency in PGS-4. The dominant Standard Model (SM) backgrounds are
\begin{eqnarray}
t\bar{t}W^\pm\to b\bar{b}W^\pm W^\pm W^\mp; \ t\bar{t}Z\to b\bar{b}W^+W^-\ell^+\ell^-.
\end{eqnarray}
Background events are generated with the Madgraph event generator~\cite{mg} and then passed to Pythia for parton shower and hadronization. The QCD corrections to the background processes have also been computed, and the
next-to-leading (NLO) $K$-factors of order 1.1 (1.3) and 1.39 (1.39) for $t\bar{t}W^\pm$~\cite{NLOttw} and $t\bar{t}Z$~\cite{NLOttz} at 8 (14) TeV LHC are used in our calculation respectively.

\begin{figure}[t]
\begin{center}
\includegraphics[scale=1,width=7cm]{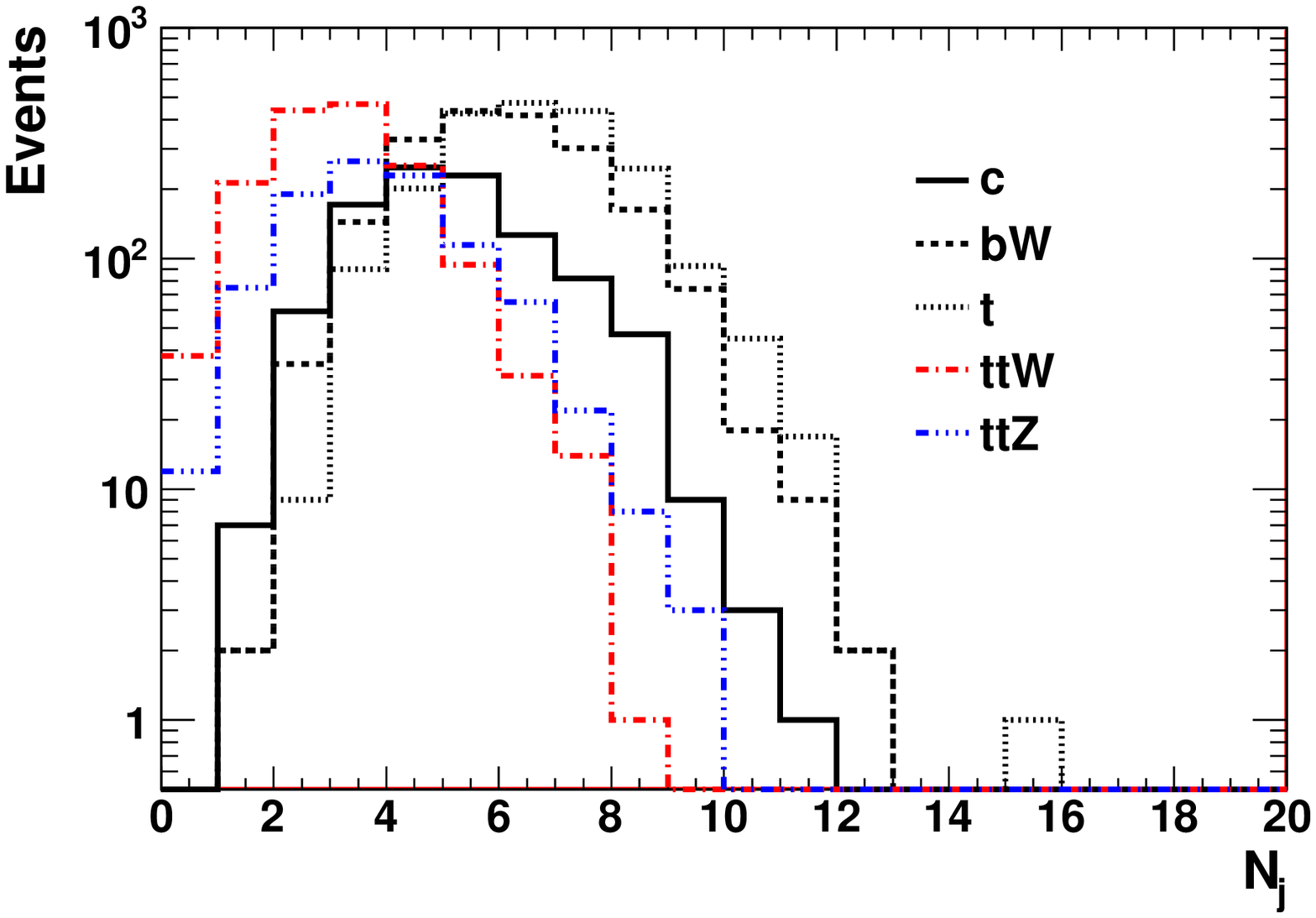}\\
\includegraphics[scale=1,width=7cm]{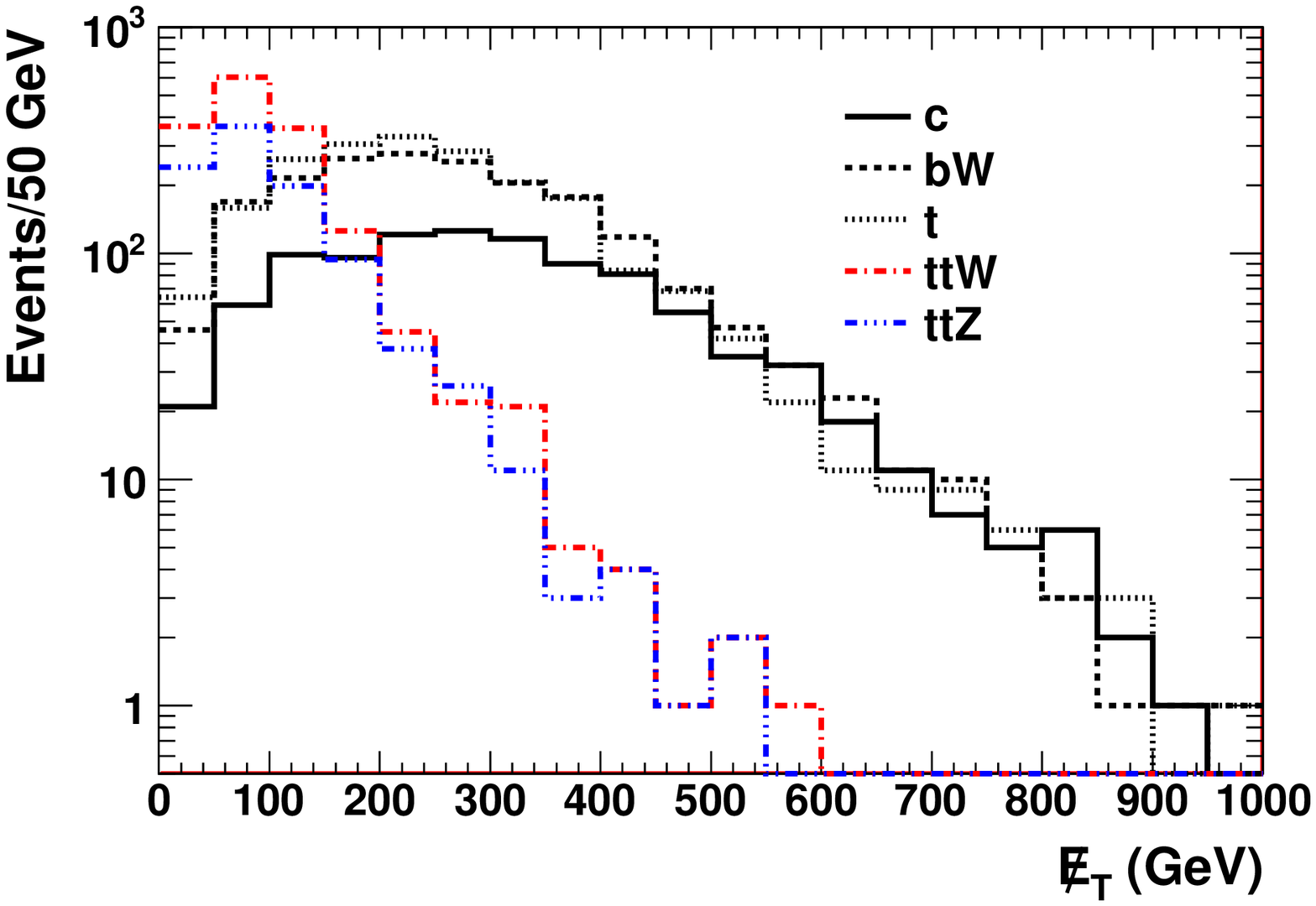}
\includegraphics[scale=1,width=7cm]{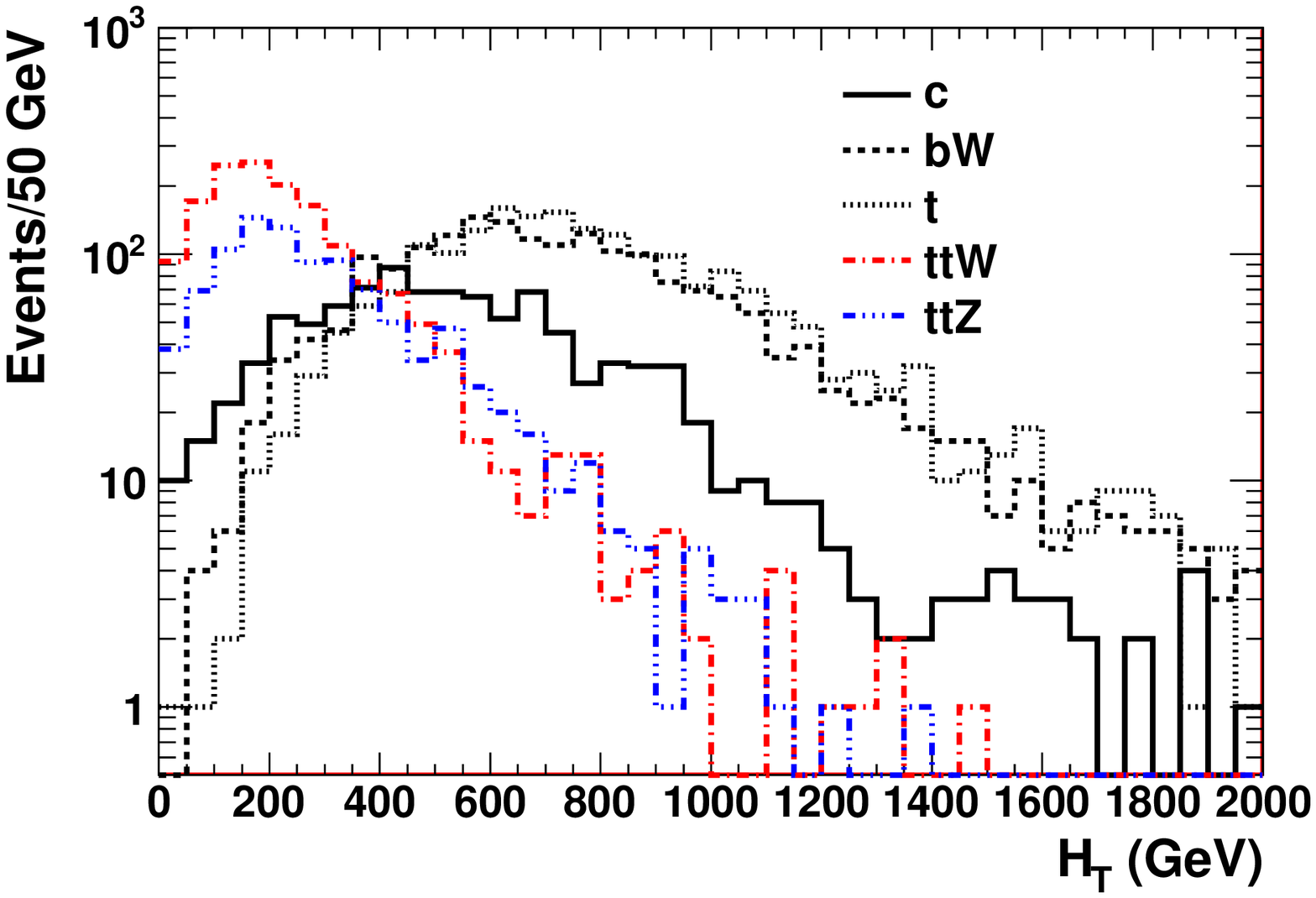}
\end{center}
\caption{Selected events with same-sign dilepton vs. number of jets $N_j$ (top), transverse missing energy $\cancel{E}_T$ (bottom left) and $H_T$ (bottom right) at 14 TeV LHC with the integrated luminosity of 100 fb$^{-1}$. The sample masses for c, bW and t modes are taken to be $(M_{\tilde{\chi}_1^0}, M_{\tilde{t}_1})=(150 \ {\rm GeV},200 \ {\rm GeV})$, $(150 \ {\rm GeV},300 \ {\rm GeV})$ and $(150 \ {\rm GeV},400 \ {\rm GeV})$ respectively.}
\label{events14}
\end{figure}

\begin{figure}[h]
\begin{center}
\includegraphics[scale=1,width=10cm]{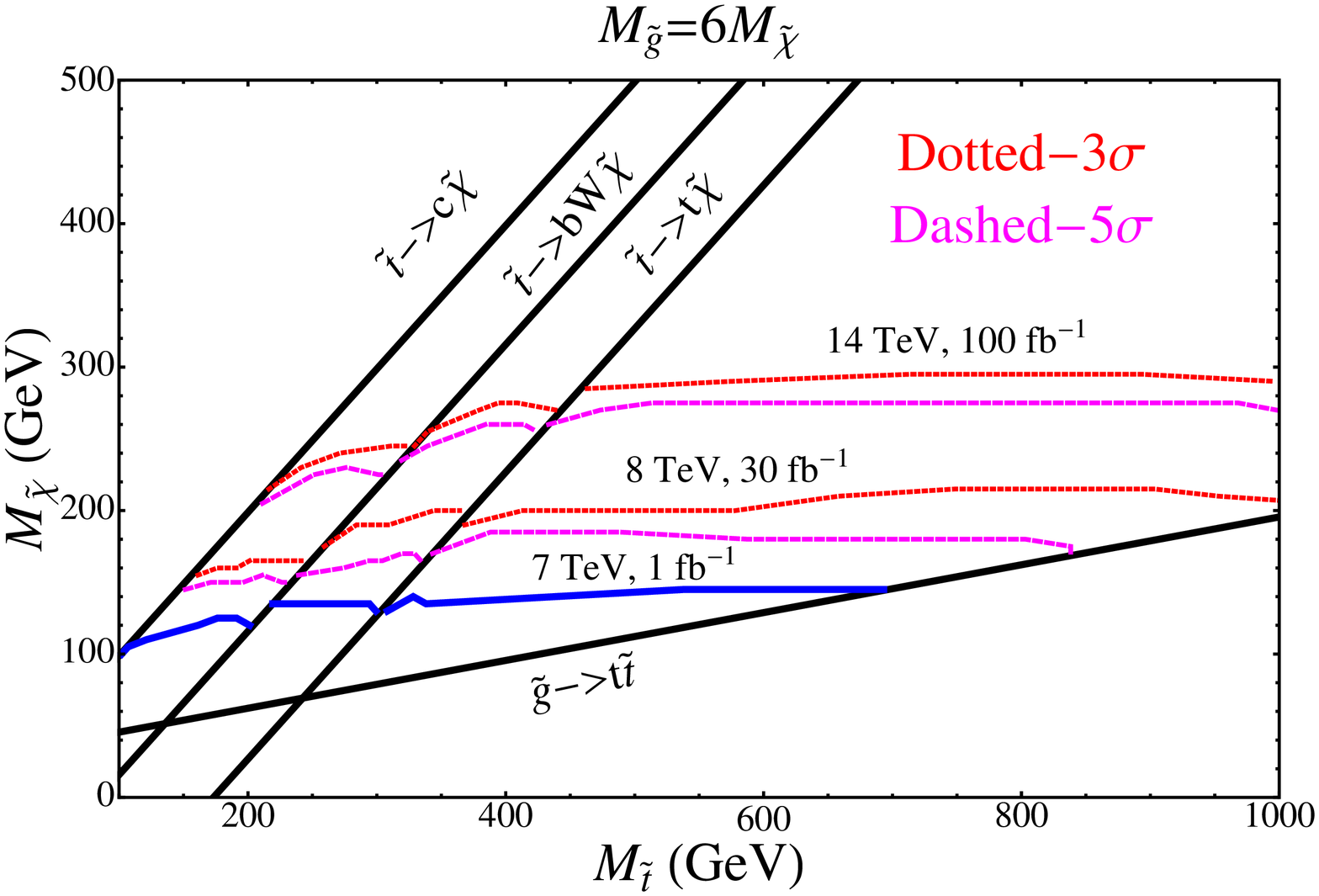}
\end{center}
\caption{Constraint and discovery limits on heavy gluino and NLSP stop scenario in the plane of $M_{\tilde{\chi}_1^0}$ vs. $M_{\tilde{t}_1}$.}
\label{significance}
\end{figure}

In Fig.~\ref{events14} we show the events of signal and background corresponding to the number of jets $N_j$ (top), transverse missing energy $\cancel{E}_T$ (bottom left) and $H_T$ (bottom right) after the above selection requirements at 14 TeV LHC with the integrated luminosity of 100 fb$^{-1}$. The sample masses for c, bW and t modes are taken to be $(M_{\tilde{\chi}_1^0}, M_{\tilde{t}_1})=(150 \ {\rm GeV},200 \ {\rm GeV})$, $(150 \ {\rm GeV},300 \ {\rm GeV})$ and $(150 \ {\rm GeV},400 \ {\rm GeV})$ respectively. We further propose some judicious cuts to suppress SM backgrounds
\begin{itemize}
\item $N_j\geq 4$, $N_b\geq 1$ for c mode, and $N_j\geq 4$, $N_b\geq 2$ for bW and t modes,
\item $\cancel{E}_T>100$ GeV for 8 TeV, and $\cancel{E}_T>150$ GeV for 14 TeV,
\item $H_T>150$ GeV for 8 TeV, and $H_T>300$ GeV for 14 TeV.
\end{itemize}
After the above kinematic cuts, in Fig.~\ref{significance}, one can see that the 8 TeV LHC with 30 fb$^{-1}$ luminosity can reach 960 (900) GeV, 1.2 (1.0) TeV and 1.3 (1.1) TeV of gluino mass for c, bW and t mode respectively, supposing $S/\sqrt{B}>3 (5) \sigma$ significance. The discovery limits of heavy gluino mass at the 14 TeV LHC with 100 fb$^{-1}$ luminosity are 1.45 (1.35) TeV, 1.6 (1.55) TeV and 1.9 (1.68) TeV for c, bW and t mode respectively with significance $S/\sqrt{B}>3 (5) \sigma$.

%%%%%%%%%%%%%%%%%%%%%

\section{Conclusions}
Motivated by the lack of evidence for low-energy SUSY from the search for light gluino and first two family squarks at 7 TeV LHC, we study the scenario with NLSP stop which is favored by dark matter accommodation, electroweak baryongenesis and the naturalness in MSSM. Considering a simplified spectrum, the light NLSP stop is typically produced associated with top quark in pair as decay products from heavier gluino pair production, namely $\tilde{g}\to t\tilde{t}^\ast_1+\bar{t}\tilde{t}_1$ with 100\% decay branching fraction. The NLSP stop $\tilde{t}_1$ is assumed to essentially decay into $c\tilde{\chi}_1^0$, $bW^+\tilde{\chi}_1^0$ or $t\tilde{\chi}_1^0$, denoted as c mode, bW mode and t mode respectively.

We have employed the CMS analysis of events with same-sign dilepton, corresponding to 1 fb$^{-1}$ of integrated luminosity, to impose constraint on this scenario. We obtain a lower bound in this scenario of 750 GeV (840 GeV) on the gluino mass for c (bW and t) mode.

After applying the same object requirements as in CMS detector and judicious cuts, we find that the 8 TeV LHC with 30 fb$^{-1}$ luminosity can reach 960 (900) GeV, 1.2 (1.0) TeV and 1.3 (1.1) TeV of gluino mass for c, bW and t mode respectively, supposing significance $S/\sqrt{B}>3 (5) \sigma$. The discovery limits of heavy gluino mass at the 14 TeV LHC with 100 fb$^{-1}$ luminosity are 1.45 (1.35) TeV, 1.6 (1.55) TeV and 1.9 (1.68) TeV for c, bW and t mode respectively with significance $S/\sqrt{B}>3 (5) \sigma$.
%%%%%%%%%%%%%%%%%%%%%%%%%%
\subsection*{Acknowledgment}
%%%%%%%%%%%%%%%%%%%%%%%%%%
This work is supported by the DOE under grant No. DE-FG02-12ER41808.

%%%%%%%%%%%%%%%%%%%%%%%%%%%%%%%%%%%%%%%%%%%%%%


\begin{thebibliography}{99}

\bibitem{cmssm}
A.~Chamseddine, R.~Arnowitt and P.~Nath, Phys.\ Rev.\ Lett.\ {\bf 49} (1982) 970; R.~Barbieri, S.~Ferrara and C.~Savoy, Phys.\ Lett.\ {\bf B119} (1982) 343;
N.~Ohta, Prog.\ Theor.\ Phys.\ {\bf 70} (1983) 542; L.~J.~Hall, J.~D.~Lykken and S.~Weinberg, Phys.\ Rev.\ {\bf D27} (1983) 2359; for a review see H.~P.~Nilles, Phys.\ Rep.\ {\bf 110} (1984) 1; S. Weinberg, {\it The Quantum Theory of Fields: Volume 3, Supersymmetry, Cambridge University Press (2000) 442p}.

\bibitem{atlasjets}
ATLAS Collaboration, ATLAS-CONF-2012-033.

\bibitem{cmsjets}
CMS Collaboration, CMS-PAS-SUS-11-016.

\bibitem{mmssm}
A.~G.~Cohen, D.~B.~Kaplan and A.~E.~Nelson, Phys.\ Lett.\ {\bf B388} (1996) 588-598.

\bibitem{lsp}
For a review see G.~Jungman, M.~Kamionkowski and K.~Griest, Phys.\ Rep.\ {\bf 267} (1996) 195.

\bibitem{wmap}
WMAP Collaboration E.~Komatsu et al., Astrophys.\ J.\ Suppl.\ {\bf 192} (2011) 18.

\bibitem{relicabundance}
S.~Profumo and C.~E.~Yaguna, Phys.\ Rev.\ {\bf D70} (2004) 095004.

\bibitem{carlos}
M.~Carena, G.~Nardini, M.~Quiros and C.~Wagner, Nucl.\ Phys.\ {\bf B812} (2009) 243.


\bibitem{naturalness}
R.~Kitano and Y. Nomura, Phys.\ Rev.\ {\bf D73} (2006) 095004; M.~Asano, H.~D.~Kim, R.~Kitano and Y.~Shimizu, JHEP {\bf 1012} (2010) 019.


%\bibitem{btau}
%I.~Gogoladze, S.~Raza and Q.~Shafi, arXiv: 1104.3566 [hep-ph].

\bibitem{expstop1}
K.~Nakamura et al. (Particle Data Group), J.\ Phys.\ G{\bf 37} (2010) 075021 and 2011 partial update for the 2012 edition.

\bibitem{expstop2}
R.~Demina, J.~D.~Lykken, K.~T.~Matchev and A.~Nomerotski, Phys.\ Rev.\ {\bf D62} (2000) 035001.

\bibitem{expstop3}
CDF Collaboration, CDF Note 9834, see http://www-cdf.fnal.gov/physics/exotic/r2a/20090709.stop\_charm/ for details.

%\bibitem{kobayashi}
%K.~I.~Hikasa and M.~Kobayashi, Phys.\ Rev.\ {\bf D36} (1987) 724; M.~Muhlleitner and E.~Popenda, JHEP {\bf 1104} (2011) 095.


\bibitem{carlos2}
M.~Carena, A.~Freitas and C.~Wagner, JHEP {\bf 10} (2008) 109.

\bibitem{adeel}
M.~A.~Ajaib, T.~Li and Q.~Shafi, Phys.\ Rev.\ {\bf D85} (2012) 055021.

\bibitem{bin}
B.~He, T.~Li and Q.~Shafi, JHEP {\bf 05} (2012) 148.

\bibitem{kraml}
S.~Kraml and A.~R.~Raklev, Phys.\ Rev.\ {\bf D73} (2006) 075002; S.~Kraml and A.~R.~Raklev, AIPConf.\ Proc.\ {\bf 903} (2007) 225.

\bibitem{martin}
S.~P.~Martin, Phys.\ Rev.\ {\bf D78} (2008) 055019.

\bibitem{atlasss2}
ATLAS Collaboration, arXiv: 1203.5763 [hep-ex].


\bibitem{cmsss5}
CMS Collaboration, CMS-PAS-SUS-11-010 and update with full 2011 data, arXiv: 1205.6615 [hep-ex].



\bibitem{simmodel1}

J.~Alwall, M.~P.~Le, M.~Lisanti and J.~Wacker, Phys.\ Lett.\ {\bf B666} (2008) 34; J.~Alwall, M.~P.~Le, M.~Lisanti and J.~Wacker, Phys.\ Rev.\ {\bf D79} (2009) 015005.

\bibitem{simmodel2}
N.~Arkani-Hamed, P.~Schuster, N.~Toro, J.~Thaler, L.~T.~Wang, B.~Knuteson and S.~Mrenna, arXiv: 0703088.

\bibitem{simmodel3}
J.~Alwall, P.~Schuster, N.~Toro, Phys.\ Rev.\ {\bf D79} (2009) 075020.

\bibitem{simmodel4}
D.~Alves et al., arXiv: 1102.5338 [hep-ph].

\bibitem{pythia}
T.~Sjostrand, S.~Mrenna and P.~Skands, JHEP {\bf 0605} (2006) 026.

\bibitem{pgs}
J.~Conway et al., http://www.physics.ucdavis.edu/$\sim$conway/research/software/pgs/pgs4-general.htm.

\bibitem{prospino}
Prospino 2.1, available at http://www.thphys.uni-heidelberg.de/$\sim$plehn/index.php?show=prospino\&visible=tools; W.~Beenakker, R.~Hopker, M.~Spira and P.~M.~Zerwas, Nucl.\ Phys.\ {\bf B492} (1997) 51; W.~Beenakker et al, Nucl.\ Phys.\ {\bf B515} (1998) 3; W.~Beenakker et al, Phys.\ Rev.\ Lett.\ {\bf 83} (1990) 3780 [Erratum-ibid. 100 (2008) 029901]; M.~Spira, hep-ph/0211145; T.~Plehn, Czech.\ J.\ Phys.\ {\bf 55} (2005) B213.

\bibitem{cmsssb5}
CMS Collaboration, CMS-PAS-SUS-11-020.

\bibitem{mg}
J.~Alwall, M.~Herquet, F.~Maltoni, O.~Mattelaer and T.~Stelzer, JHEP {\bf 1106} (2011) 128.


\bibitem{NLOttw}
J.~M.~Campbell and R.~K.~Ellis, arXiv: 1204.5678 [hep-ph].

\bibitem{NLOttz}
A.~Kardos, Z.~Trocsanyi and C.~Papadopoulos, arXiv: 1111.0610 [hep-ph].

\end{thebibliography}
\end{document}